# Sub-Atomic Channeling and Vortex Beams in SrTiO$_3$

Jong Seok Jeong*, Hosup Song, Jacob T. Held, K. Andre Mkhoyan*

Department of Chemical Engineering and Materials Science, University of Minnesota, Minneapolis, Minnesota 55455, United States

*Corresponding Authors: jsjeong@umn.edu (JSJ), mkhoyan@umn.edu (KAM)

**Abstract:**

Inspired by recent experimental sub-atomic measurements using analytical aberration-corrected scanning transmission electron microscopes (STEMs), we studied electron probe propagation in crystalline SrTiO$_3$ at the sub-atomic length scale. Here, we report the existence of sub-atomic channeling and the formation of a vortex beam at this scale. The results of beam propagation simulations, which are performed at various crystal temperatures and STEM probe convergence angles (10 to 50 mrad) and beam energies (80 to 300 keV), showed that reducing the ambient temperature can enhance the sub-atomic channeling and that STEM probe parameters can be used to control the vortex beams and adjust their intensity, speed, and area.

Aberration correction of the lenses in the transmission electron microscope (TEM) has allowed the field of analytical microscopy [1-3], particularly analytical scanning TEM (STEM), to enter a new realm of resolution. As a result, it is possible to detect atomic columns in crystals with < 1 Å separation and even individual atoms [4,5]. The development of atomic-resolution spectroscopy has naturally followed that of atomic-scale imaging [6-11]. Yet, these instruments can be pushed further and can offer a window into sub-atomic physics via measurements that previously were not considered possible. A report by Muller et al. [12] showed that by using aberration-corrected STEM, it is possible to measure sub-atomic electrical fields inside crystals. In another example, the report by Jeong et al. [13] showed that it is possible to probe core electronic orbitals of atoms and, in the process, to measure the impact parameter for electronic excitations from core levels to higher levels.

One way to approach sub-atomic STEM measurements is to explore the propagation of an aberration-corrected STEM probe through crystalline samples at the sub-atomic level. Early studies of beam propagation through crystals at the atomic level revealed many interesting phenomena, including beam channeling through the atomic columns in crystals [14-16] and beam intensity transfer from one atomic column to another [17]. Understanding beam channeling was instrumental for quantification of atomic-resolution annular dark-field (ADF) STEM images [18,19], which, in turn, aided in the quantification of individual dopant atoms in crystalline materials [20,21].

In this letter, we present a simulation-based study of sub-atomic channeling of the STEM probe through a $SrTiO_3$ crystal, which, as will become clear, is already rich with new phenomena, from sub-atomic pendulum-like oscillations of the beam to the formation of vortex beams. The implications of these observations are also discussed.

To investigate sub-atomic channeling, a SrTiO$_3$ crystal was selected as a model system. Studying channeling in SrTiO$_3$ along the [001] crystallographic orientation has several advantages: it is well studied both experimentally and theoretically and, therefore, parameters are well known [22]; it has large spacing between atomic columns ($a = 3.905$ Å), which simplifies the visualization of the beam channeling; and it also has light (O) and heavy (Sr) atomic columns, and also the combination of both (Ti/O). Fig. 1(a) shows a schematic of a SrTiO$_3$ unit cell viewed along the [001] crystallographic orientation. It also shows several probe locations that were used to study beam propagation in SrTiO$_3$. For the unit cell of SrTiO$_3$ in the [001] projection, a triangle connecting the Sr, Ti/O, and O columns is the smallest sampling area with unique STEM probe locations [Fig. 1(a)].

First, the propagation of the aberration-corrected STEM probe in SrTiO$_3$ was studied as a function of the distance of the probe from the atomic column at sub-atomic distances. This was simulated using the Multislice method [23] as implemented by the TEMSIM code [24]. To ensure accurate thermal diffuse scattering (TDS)-inclusive simulations at room temperature (T = 300 K), RMS thermal vibration values were determined from the experimental diffraction literature: 0.0773 Å$^2$ for Sr, 0.0606 Å$^2$ for Ti, and 0.0848 Å$^2$ for O [13,25]. For these calculations, an aberration-free STEM probe was used that had a convergence angle of $\alpha_{obj} = 24.5$ mrad and beam energy of $E = 300$ keV [16]. Fig. 1 shows the results of the beam propagation when the STEM probe location was gradually moved from the Sr atomic column to a neighboring Ti/O column (for beam propagation with finer probe steps see SI, video S1).

As can be seen from Fig. 1, when the probe is located exactly on top of the atomic columns, for instance Sr (position *P1*) and Ti/O (position *P9*), the expected on-column channeling is observed [16]. However, when the probe is slightly moved away from the atomic column, at a distance smaller than the size of the atomic dimensions, the on-column channeling disappears. Instead, the propagating beam oscillates back and forth around the center of the atomic column, like a pendulum. Additionally, the frequency of these oscillations is different than that for the on-column channeling. These unique oscillations can be observed even when the probe is located only 0.2 Å away from the column. It appears that when the probe is located approximately 0.4 Å away from the Sr or Ti/O columns, the oscillations are most prominent, and then they die off, reconfirming the sub-atomic nature of the oscillations. For simplicity of discussion, we will refer to these sub-atomic, pendulum-like oscillations as 'sub-atomic' channeling.

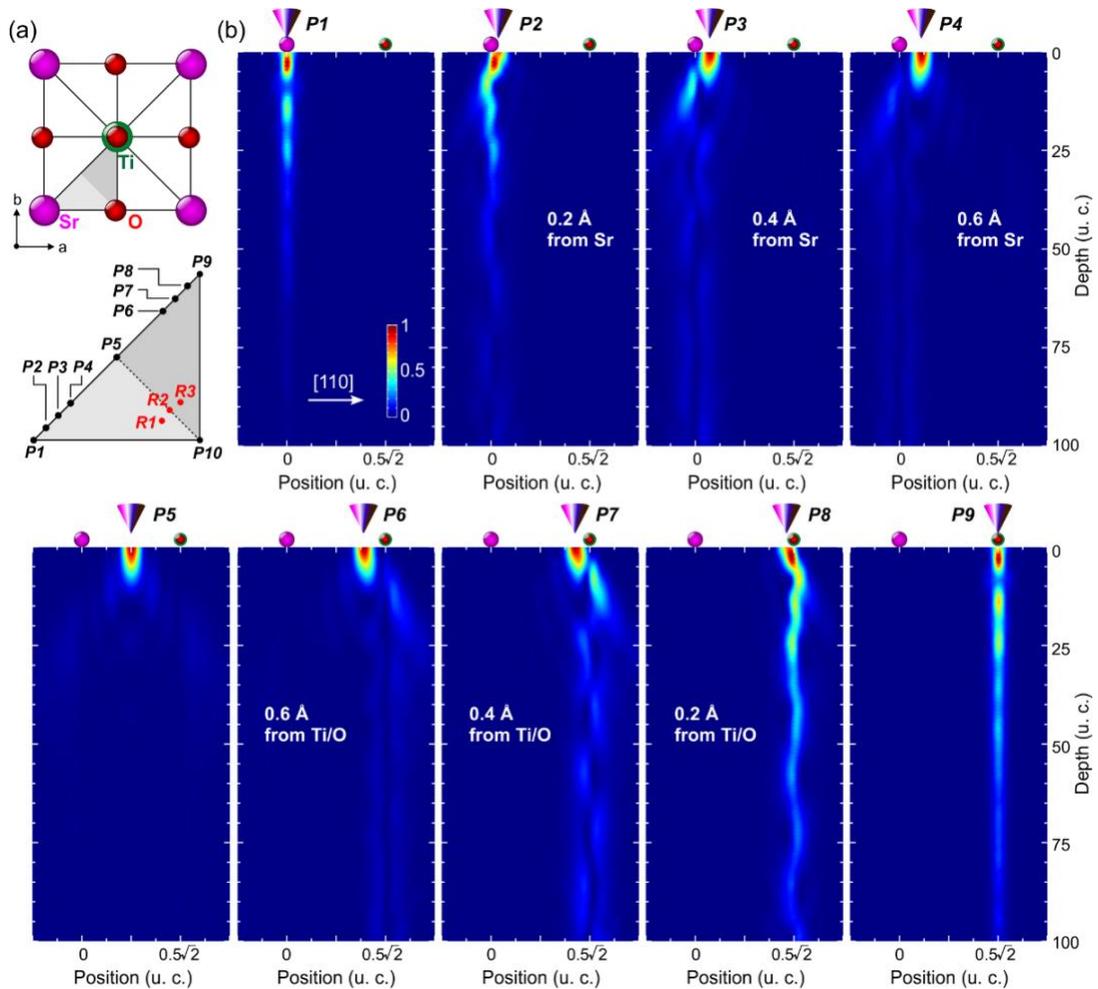

**Figure 1.** (a) Schematic illustration of a SrTiO$_3$ unit cell viewed along the [001] crystallographic orientation. The gray-colored triangle connecting Sr, Ti/O, and O atomic columns represents the minimum symmetric unit with unique probe positions. STEM probe positions (*P1-P10 and R1-R3*) studied in this work are also shown. (b) Two-dimensional (2-D) beam intensity depth profiles of a simulated STEM probe when the probe moves from Sr to Ti/O atomic columns. These cross-sections are along the [110] direction. The atomic column positions are indicated by the colored circles above the panels, and the incident probe positions are indicated by the inverted cones. The sampling step is 0.2 Å with one additional probe location at the middle of two columns. The size of the unit cell in SrTiO$_3$ is $a$ = 3.905 Å.

The results of the simulations presented in Fig. 1 (and also in the SI) capture several unique characteristics of this sub-atomic channeling: (1) the frequency of the oscillations is a function of the distance of the initial probe position relative to the center of the atomic column, (2) for the probes located at the same distance from the columns, the oscillations are stronger for the column with heavy atoms, and (3) the propagating beam, while it oscillates within the dimensions of the atom, avoids being in the center of the atom, which is in complete contrast to on-column channeling.

The sensitivity of this sub-atomic channeling to crystal temperature and STEM probe parameters was also tested. Lowering the temperature of the crystal results in a reduction of the thermal vibrations of the atoms, which in turn reduces the TDS of the probe electrons [26]. Fig. 2 shows the results of simulation for beam propagation when the lattice phonons are quenched to experimentally achievable liquid nitrogen temperature, T = 77 K. At lower temperatures, the sub-atomic channeling clearly becomes enhanced (see Fig. 1(b) for comparison). The simulations, additionally performed at various electron beam energies and probe convergence angles, showed that the sub-atomic channeling can be observed at least for the energy range of 80 to 300 keV and for convergence angles of 10 to 50 mrad (see SI).

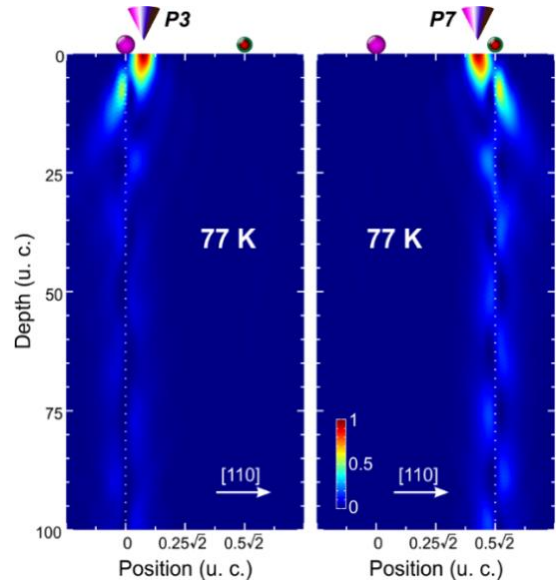

**Figure 2**. 2-D beam intensity depth profiles of a simulated STEM probe when the probe is located at 0.4 Å from the Sr and Ti/O atomic columns. The simulations were performed for the SrTiO$_3$ crystal at 77 K. These cross-sections are comparable to the results for probe positions *P3* and *P7* in Fig. 1(b).

When the probe occupies certain locations, i.e., very close to the column of O atoms inside the triangle connecting the Sr, Ti/O, and O columns, the propagating probe electrons form a vortex beam [27], circulating around the O column. By selecting the location of the incident probe within the triangle, both left-handed and right-handed vortex beams can be obtained: that is, the chirality of the vortex beam depends on where the probe located, either close to a Sr column (lighter shaded region) or a Ti/O column (darker shaded region). Two propagating beams from representative probe positions in these regions, forming left-handed and right-handed vortex beams, are shown in Fig. 3 (for full beam propagation with vortex beam formation, see SI, video S2). Interestingly, when the probe is located at the border of the two regions, the vortex beam disappears and oscillatory sub-atomic channeling reappears [Fig. 3(b)].

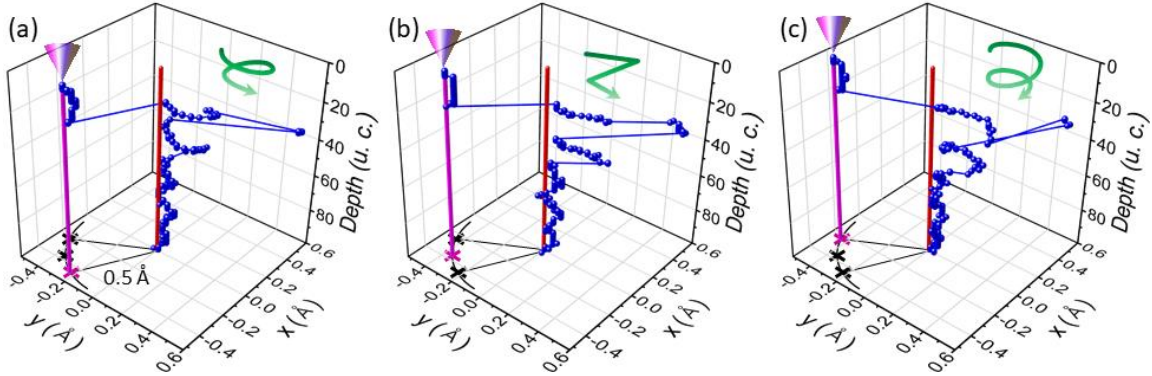

**Figure 3.** The formation of a vortex beam around the O atomic column in the SrTiO$_3$ crystal. The paths of the electron beams are represented by the maximum intensity positions for the electron probe located at positions *R1* (a), *R2* (b), and *R3* (c) [see Fig. 1(a)]. In all three cases, the probe is located 0.5 Å away from the O atomic column. The *R1* and *R3* are located slightly off the border line, which connects two close O columns in the unit cell, and the *R2* is located on that line. The probe positions and O atomic column are indicated by the magenta and red vertical lines, respectively. Positions *R1* and *R3* produce vortex beams with opposite chirality, and position *R2* produces an oscillating beam.

These vortex beam formations can be described by considering the atomic potentials around the STEM probe and the forces that they exert on the electrons of the probe inside the crystal. Such a description is schematically demonstrated in Fig. 4(a). When the probe is located very close to the O column and relatively close to the Sr column, a portion of the probe electrons, shown by the dotted line, is pulled away by the Sr atoms, and the resulting non-circularly symmetric probe is first pulled toward the O column and then spirals within the potential of the O atom. This imbalance of electrons in the probe is shifted to the opposite direction when the probe is located closer to a Ti/O column, and the resulting propagating vortex beam possesses opposite chirality. The vortex beam gradually focuses as it propagates through the thickness of the crystal; however, it can be visible even at depths of approximately 80-100 unit cells (or 310-390 Å).

The formation of vortex beam can be observed in reciprocal space as well. Fig. 4(b) shows the results of simulation for beam propagating through SrTiO$_3$ in reciprocal space with probe located at positions *R1* and *R3* (for full beam propagation, see SI, video S3). Already, at the depth of 36 u.c. (or 14.0 nm), rotation of the beam in counter clockwise for position *R1* and clockwise for position *R3* is observable. Such opposite rotation of the beam in reciprocal space is a result of an uncompensated beam with average wave vector having non-zero $k_{xy}$ component, which results in formation of left-handed and right-handed vortex beams.

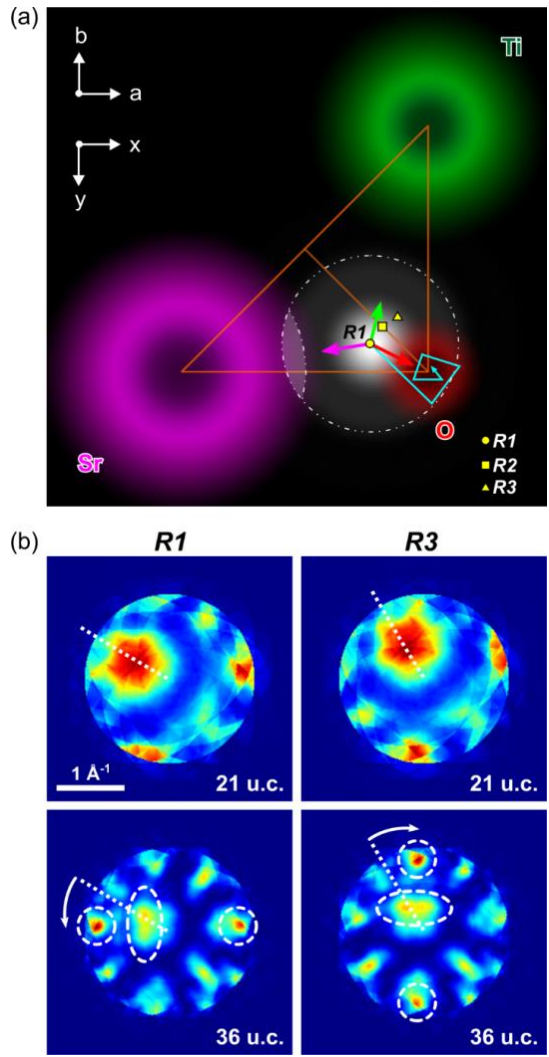

**Figure 4.** (a) Illustration describing formation of the vortex beam inside SrTiO$_3$. The incident STEM probe located at the point *R1* and the atomic potentials (Sr: magenta; Ti/O: green; and O: red). The triangle connecting the Sr, Ti/O, and O atomic columns are indicated [see also Fig. 1(a)]. The dotted circle highlights the simulated incident probe. The section of the probe within the potential of the Sr atom is shaded within the dotted circle. The electron beam path as it propagates through SrTiO$_3$ is indicated by the blue line, illustrating the formation of a vortex beam around the O atomic column. The probe positions *R1*, *R2*, and *R3* discussed in Fig. 3 are indicated. The crystallographic orientation and the *x* and *y* axes are also shown. Coulombic forces from all three atomic columns attracting the propagating electron beam are represented by the arrows with the same colors of the atomic columns. (b) The intensity of propagating beams in reciprocal space, located at positions *R1* and *R3*, at two different depths showing left-handed and right-handed rotations, respectively.

To evaluate the properties of these vortex beams and their dependence on the STEM probe parameters, additional simulations were performed for the probes with various convergence angles and beam energies. The results show that with an increase in the probe convergence angle, from $\alpha_{obj} = 10$ mrad to 50 mrad, the speed of the beam rotation can be considerably increased (see SI, video S4). However, a reduction in the beam energy from 300 keV to 80 keV only dampened the rotation of the beam (see SI, video S5), suggesting that for enhanced chirality of the propagating beam, higher beam energies as well as higher probe convergence angles are desirable.

In conclusion, we have shown that there exists rich sub-atomic physics to be explored using available aberration-corrected analytical STEMs. To do so, understanding the propagation of the STEM probe through a crystalline material at the sub-atomic length scale is essential. The importance of this is highlighted here by demonstrating an existence of sub-atomic channeling and a formation of vortex beams in a $SrTiO_3$ crystal. We also demonstrated that the ambient temperature can influence this sub-atomic channeling, and that the STEM probe parameters can be used to control the vortex beams. In addition to the possibility to exploit sub-atomic channeling to probe atomic electronic orbitals, the vortex beams can provide a window into measurements of magnetism from individual atomic columns.

**Acknowledgments:**

This work was supported in part by the NSF MRSEC under award number DMR-1420013, and in part by the Grant-in-Aid program of the University of Minnesota. The channeling simulations were performed at the Minnesota Supercomputing Institute at the University of Minnesota.

**Supplemental Material:**

See Supplemental Material for the additional beam intensity depth profiles of a simulated STEM probe when the probe moves from Sr atomic columns to O and Ti/O atomic columns. The videos show full three-dimensional (3-D) beam propagation through the $SrTiO_3$ crystal and the formation of vortex beam simulated for various STEM probe convergence angles and beam energies.